\newcommand*{\RELEASE}{}  
\newcommand*{\NOCOPYRIGHT}{}  

\documentclass[conference]{IEEEtran}
\IEEEoverridecommandlockouts
\usepackage{cite}
\usepackage{amsmath,amssymb,amsfonts}
\usepackage{algorithmicx}
\usepackage{graphicx}
\usepackage{textcomp}
\def\BibTeX{{\rm B\kern-.05em{\sc i\kern-.025em b}\kern-.08em
    T\kern-.1667em\lower.7ex\hbox{E}\kern-.125emX}}

\graphicspath{{fig/}{../fig/}}

\usepackage[utf8]{inputenc}
\usepackage{paralist}  
\usepackage{balance}
\usepackage{xspace}  
\usepackage{setspace}  
\usepackage{url}

\usepackage{booktabs}  
\usepackage{multirow}  
\usepackage{stfloats}  

\usepackage{subfigure}  
\usepackage{wrapfig}  
\usepackage{floatflt}  

\usepackage{algorithm}
\usepackage{algpseudocode}  

\usepackage[table]{xcolor}  

\definecolor{green(pigment)}{rgb}{0.0, 0.65, 0.31}

\ifdefined\RELEASE  
	\newcommand{\al}[1]{} 
	\newcommand{\refine}[2]{#1}
	\newcommand{\del}[1]{}  

	\newcommand{\mkh}[1]{}
	\newcommand{\jgrm}[1]{}
\else
	\newcommand{\al}[1]{\textcolor{green(pigment)}{[AL: #1]}} 
	\newcommand{\refine}[2]{\textcolor{violet}{#1}\textcolor{teal}{[#2]}}
	\newcommand{\del}[1]{\textcolor{blue}{\sout{#1}}}  

	\newcommand{\mkh}[1]{\textcolor{brown}{(*** MK: #1 ***)}}
	\newcommand{\jgrm}[1]{\textcolor{brown}{\sout{#1}}}
\fi

\newcommand{\anm}{ANONYMIZED\xspace}
\ifdefined\ANONYMOUS
	\newcommand{\urlSys}{\anm} 
	\newcommand{\urlBmark}{\anm}  
	\newcommand{\citeStx}{\anm} 
\else
	\newcommand{\urlSys}{https://github.com/eXascaleInfolab/daoc}
	
	\newcommand{\urlBmark}{\url{https://github.com/eXascaleInfolab/clubmark}\xspace}  

	\newcommand{\citeStx}{StaTIX~\cite{Stx18}} 
\fi

\ifdefined\NOCOPYRIGHT
	\newcommand{\IEEEcpr}{} 
\else
	\newcommand{\IEEEcpr}{978-1-7281-0858-2/19/\$31.00 \copyright~2019 IEEE}
\fi

\newcommand{\sys}{DAOC\xspace}
\newcommand{\bmark}{Clubmark\xspace}

\newtheorem{definition}{Definition}  
\newtheorem{example}{Example}

\newcommand\blfootnote[1]{%
  \begingroup
  \renewcommand\thefootnote{}\footnote{#1}%
  \addtocounter{footnote}{-1}%
  \endgroup
}

\begin{document}

\setstretch{0.982}  

\title{DAOC: Stable Clustering of Large Networks
}

\ifdefined\ANONYMOUS
\author{\IEEEauthorblockN{\anm}
\\
\\
}
\else
\author{\IEEEauthorblockN{Artem Lutov}  
\IEEEauthorblockA{\textit{eXascale Infolab}\\
\textit{University of Fribourg}\\
Switzerland\\
artem.lutov@unifr.ch}
\and
\IEEEauthorblockN{Mourad Khayati}  
\IEEEauthorblockA{\textit{eXascale Infolab}\\
\textit{University of Fribourg}\\
Switzerland\\
mourad.khayati@unifr.ch}
\and
\IEEEauthorblockN{Philippe Cudr{\'e}-Mauroux}  
\IEEEauthorblockA{\textit{eXascale Infolab}\\
\textit{University of Fribourg}\\
Switzerland\\
pcm@unifr.ch}
}
\fi

\maketitle

\begin{abstract}
Clustering is a crucial component of many data mining systems involving the analysis and exploration of various 
data. Data diversity calls for clustering algorithms 
to be accurate while providing stable (i.e., deterministic and robust) results on arbitrary input networks. Moreover, modern systems often operate on large datasets, which implicitly constrains the complexity of the clustering algorithm. Existing clustering techniques are only partially stable, however, as they guarantee either determinism or robustness. To address this issue, we introduce DAOC, a Deterministic and Agglomerative Overlapping Clustering algorithm. DAOC leverages a new technique called Overlap Decomposition to identify fine-grained clusters in a deterministic way capturing multiple optima.
In addition, it leverages a novel consensus approach, Mutual Maximal Gain, to ensure robustness and further improve the stability of the results while still being capable of identifying micro-scale clusters. Our empirical results on both synthetic and real-world networks show that DAOC yields stable
clusters while being 
on average 25\% more accurate than state-of-the-art deterministic algorithms 
without requiring any tuning.
Our approach has the ambition to greatly simplify and speed up data analysis tasks involving iterative processing (need for determinism) as well as data fluctuations (need for robustness) and to provide accurate and reproducible results.
\end{abstract}

\begin{IEEEkeywords}
stable clustering, deterministic overlapping clustering, community structure discovery, parameter-free community detection, cluster analysis.
\end{IEEEkeywords}

\blfootnote{\IEEEcpr}

\section{Introduction}
\label{sec:intro}

Clustering is a fundamental part of data mining with a wide applicability to statistical analysis and exploration of physical, social, biological and information systems.
Modeling and analyzing such systems often involves processing large complex networks~\cite{Brb02}.
Clustering 
large networks is intricate in practice, and should ideally
provide \emph{stable} results in an efficient way
in order to make the process easier for the data scientist.

Stability is pivotal for many data mining tasks since it allows to better understand whether the results are caused by the evolving structure of the network, by evolving node ids (updated labels, coordinates shift or nodes reordering), or by some fluctuations in the application of non-deterministic algorithms.
Stability\index{stability} of the results involves both determinism and robustness.
We refer to the term \emph{deterministic}\index{deterministic} in the strictest sense denoting algorithms that
\begin{inparaenum}[\itshape a\upshape)]
\item do not involve any stochastic operations and
\item produce results 
invariant of the nodes processing order.
\end{inparaenum}
\emph{Robustness} ensures that clustering results gracefully evolve with
small perturbations or changes in the input network~\cite{Krr08}.
It prevents sudden changes in the output for dynamic networks and provides the ability to tolerate noise and outliers in the input data~\cite{Rjs97}. 

Clustering a network is usually not a one-off project but 
an iterative process, where the results are visually explored and refined multiple times. The visual exploration of large networks requires to consider the specificities of human perception~\cite{Brk14,Mlr56} 
which is good at handling \emph{fine-grained hierarchies} of clusters. In addition, those hierarchies should be stable across iterations such that the user can compare previous results with new results. 
This calls for results that are both stable and fine-grained.

In this paper,
we introduce a novel clustering method called \sys to address the aforementioned issues. To the best of our knowledge, \sys\footnote{\urlSys} is the first parameter-free clustering algorithm that is simultaneously deterministic, robust and applicable to large weighted networks yielding a fine-grained hierarchy of overlapping clusters.  
%
More specifically, \sys leverages
\begin{inparaenum}[\itshape a\upshape)]
\item a novel consensus technique we call \emph{Mutual Maximal Gain (MMG)} 
to perform a robust and deterministic identification of node membership in the  clusters, and
\item a new technique for \emph{overlap decomposition (OD)} 
to form fine-grained clusters 
in a deterministic way, even when the optimization function yields a set of structurally different but numerically equivalent optima
(see \emph{degeneracy} in Section~\ref{sec:prelim}).
\end{inparaenum}
We empirically evaluate the stability of the resulting clusters produced by our approach, as well as its efficiency and effectiveness on both synthetic and real-world networks. We show that \sys yields stable clusters while being on average 25\% more accurate than state-of-the-art deterministic clustering algorithms
and more efficient than state-of-the-art overlapping clustering algorithms
without requiring any manual tuning.
In addition, we show that \sys returns on average more accurate results than any state-of-the-art clustering algorithm on complex real-world networks (e.g., networks with overlapping and nested clusters).
We foresee \sys to represent an important step forward for clustering algorithms as:
\begin{inparaenum}[\itshape a\upshape)]
\item deterministic clustering algorithms are usually not robust and have a lower accuracy than their stochastic counterparts, and
\item robust methods are typically not deterministic and do not provide fine-grained results as they are insensitive to micro-scale changes, as described in more detail in the following section. 
\end{inparaenum}



\section{Related Work}
\label{sec:relwork}

A great diversity of clustering algorithms can be found in the literature.
Below, we give an overview of prior methods achieving robust results, before describing  deterministic approaches and outlining a few widely used algorithms that are neither robust nor deterministic but were inspirational for our method.

\paragraph{Robust clustering algorithms} typically leverage \textit{consensus} or \textit{ensemble} techniques~\cite{Frd03,VgP11,Lnc12,Mdl18}. 
They identify clusters using consensus functions (e.g., majority voting) by processing an input network multiple times and varying either the parameters of the algorithm, or the clustering algorithm itself.
However, such algorithms typically
\begin{inparaenum}[\itshape a\upshape)]
\item are unable to detect fine-grained structures due to the lack of consensus therein,
\item are stochastic and
\item are inapplicable to large networks due to their high computational cost.
\end{inparaenum}
We describe some 
prominent and scalable consensus clustering algorithms 
below.

\noindent - \emph{Order Statistics Local Optimization Method (OSLOM)}~\cite{Lcn11} is one of the first widely used consensus clustering algorithms, which accounts for weights of the network links and yields overlapping clusters with a hierarchical structure. It is based on the local optimization of a fitness function expressing the statistical significance of clusters with respect to random fluctuations. 
OSLOM scales near linearly on sparse networks but has a relatively high computational complexity at each iteration, 
making it inapplicable to large real-world networks (as we show in Section~\ref{sec:evals}). 

\noindent - \emph{Core Groups Graph Clustering Randomized Greedy (CGGC[i]\_RG)}~\cite{Ogn13} is a 
fast and accurate ensemble clustering algorithm. 
It applies a generic procedure of ensemble learning called Core Groups Graph Clustering (CGGC) to determine several weak graph (network) clusterings 
and then to
form a strong clustering from their maximal overlap.
The algorithm has a near linear 
computational complexity with the number of edges due to the sampling and local optimization strategies applied at each iteration.
However, this algorithm is designed for unweighted graphs and produces flat and non-overlapping clusters only, which limits its applicability and yields low accuracy on large complex networks as we show in Section~\ref{sec:evals}. 

\noindent - \emph{Fast Consensus} technique 
was recently proposed and works on top of state-of-the-art clustering algorithms including Louvain (FCoLouv), Label Propagation (FCoLPM) and Infomap (FCoIMap)~\cite{Adt19}. 
The technique initializes a consensus matrix and then iteratively refines it until convergence as follows. First, the input network is clustered by the original algorithm multiple times. The consensus values $D_{i,j} \in [0, 1]$ of the matrix are evaluated as the fraction of the runs in which nodes $i$ and $j$ belong to the same cluster. The consensus matrix is formed using pairs of co-clustered adjacent nodes and extended with closed triads instead of all nodes in the produced clusters, which significantly reduces the amount of computation.
The formed matrix is filtered with a threshold $\tau$ and then clustered $n_p$ 
times by the original clustering algorithm, producing a refined consensus matrix. This refinement process is repeated until all runs produce identical clusters (i.e.,until all values in the consensus matrix are either zero and one) with precision $1 - \delta$.
The Fast Consensus technique however lacks a convergence guarantee and relies on three parameters having a strong impact on its computational complexity.

\paragraph{Deterministic clustering algorithms} and, in general, non-stochastic ones
(i.e., algorithms relaxing the determinism constraint) 
are typically 
not robust and are sensitive to both
\begin{inparaenum}[\itshape a\upshape)]
\item initialization~\cite{SuT07,Clb15,Tpr11,Hou17} (including the order in which the nodes are processed) and
\item minor changes in the input network,
\end{inparaenum}
which may 
significantly affect the clustering results~\cite{Rjs97,Lnc12}. Non-stochastic algorithms also often yield less precise results 
getting stuck on the same local optimum until the input is updated. Multiple local optima often exist due to the \emph{degeneracy} phenomenon, 
which is explained in Section~\ref{sec:prelim} and has to be specifically addressed to create deterministic clustering algorithms that are both robust and accurate.
We describe below some of the well-known deterministic algorithms.

\noindent - \emph{Clique Percolation method (CPM)}~\cite{Drn05} 
is probably the first deterministic clustering algorithm supporting overlapping clusters and capable of providing fine-grained results. \emph{Sequential algorithm for fast clique percolation (SCP)}~\cite{Kpl08} is a CPM-based algorithm, which detects $k$-clique clusters in a single run and produces a dendrogram of clusters.
SCP produces deterministic and overlapping clusters at various scales, and shows a linear dependency of the computational complexity with the number of k-cliques in the network. 
However, SCP relies on a number of parameters 
and has an exponential worst case complexity in dense networks, 
which significantly limits its practical applicability. 

\noindent - \emph{pSCAN}~\cite{Chn16} is a fast overlapping clustering algorithm for ``exact structural graph clustering'' (i.e., it is deterministic and input-order independent). 
First, it identifies core graph vertices (network nodes), which always belong to 
exactly one cluster, forming initially disjoint clusters.
The remaining nodes are then assigned to the initial clusters, 
yielding overlapping clusters.
pSCAN relies on two input parameters, $0 < \epsilon \le 1$ and $\mu \ge 2$. The results it produces are very sensitive to those 
parameters, whose optimal values are hard to \refine{guess}{tune} for arbitrary input networks.

\paragraph{Inspirational algorithms for our method} 
%
~ \\ 
\noindent - \emph{Louvain}~\cite{Bld08} is a commonly used clustering algorithm that performs modularity optimization using a local search technique on multiple levels to coarsen clusters. It introduces modularity gain as an optimization function. The algorithm is parameter-free, returns a hierarchy of clusters, and has a near-linear runtime complexity with the number of network links.
However,
the resulting clusters are not stable 
and depend on the order in which the nodes are processed.
Similarly to Louvain, our method is a greedy agglomerative clustering algorithm, which uses modularity gain as optimization function. However, the clusters formation process in \sys 
differs a lot, addressing
the aforementioned issues of the Louvain algorithm.

\noindent - \emph{DBSCAN}~\cite{Est96} is a density-based clustering algorithms suitable to process data with noise. 
It regroups points that are close in space 
given the maximal distance between the points $\varepsilon$ and the minimal number of points $MinPts$ within an area.
DBSCAN 
is limited in discovering a large variety of clusters because of its reliance to a density parameter. It has a strong dependency on input parameters, and lacks a principled way to determine optimal values for these parameters~\cite{Chn18}. We adopt however the DBSCAN idea of clusters formation based on the extension of the densest region to prevent early coarsening and to produce a fine-grained hierarchical structure.

\section{Preliminaries}  
\label{sec:prelim}

\begin{table*}[bp]
\centering
\caption{Notations}
\label{tbl:nots}
\vspace{-8pt}
\rowcolors{1}{gray!25}{white}
\begin{tabular}{l|l}
\hline
{\it \verb|#i|}   	& {Node \verb|<i>| of the network (graph)  $\mathcal{G}$}
\\ 
\textit{Q}				& \emph{Modularity}
\\ 
$\Delta Q_{i,j}$	& \emph{Modularity Gain} between \verb|#i| and \verb|#j|
\\ 
\multirow{1}{*}{$j \sim i$}
				& Items $i$ and $j$ (nodes or clusters) are neighbors (adjacent, linked)
\\ 
\multirow{1}{*}{$\Delta Q_i$}
			& Maximal \textit{Modularity Gain} for \verb|#i|:
			$\Delta Q_i = \{\max{\Delta Q_{i,j}} \;|\; \verb|#j| \sim \verb|#i|\}$
\\ 
\multirow{1}{*}{$MMG_{i,j}$}
				& \emph{Mutual Maximal Gain}:
				$MMG_{i,j} = \{\Delta Q_{i,j} \;|\; \Delta Q_i = \Delta Q_j, \: \verb|#j| \sim \verb|#i|\}$
\\ 
$ccs_i$			& Mutual clustering candidates of \verb|#i| (by $MMG_i$)
\\ \hline
\end{tabular}
\vspace{-4pt}
\end{table*}
A clustering algorithm is applied to a network to produce groups of nodes that are called \emph{clusters}\index{cluster} (also known as communities, modules, partitions or covers). Clusters represent groups of tightly-coupled nodes with loosely inter-group connections~\cite{Nwm03}, where the group structure is defined by the clustering optimization function.
The resulting clusters can be overlapping, 
which happens in case they share some common nodes, the \emph{overlap}\index{overlap}.  
The input network (graph) can be weighted and directed, where a node (vertex) weight is represented as a weighted link (edge) to the node itself (a self-loop).
The main notations used in 
this paper are listed in Table~\ref{tbl:nots}.

Clustering algorithms can be classified by the kind of input data they operate on:
\begin{inparaenum}[\itshape a\upshape)]
\item graphs specified by pairwise relations (networks) or
\item attributed graphs (e.g., vertices specified by coordinates in a multi-dimensional space). 
\end{inparaenum}
These two types of input data cannot be unambiguously converted into each other, at least unless one agrees on some customized and specific conversion function. 
Hence, their respective clustering algorithms are not (directly) comparable. In this paper, we focus on clustering algorithms working on graphs specified by pairwise relations (networks), which are also known as community structure discovery algorithms.


\paragraph{Modularity ($Q$)~\cite{Nwm04u}} is a standard measure of clustering quality 
that is equal to the difference between the density of the links in the clusters and \refine{the}{their} expected density: 
\begin{align}  
Q &= \frac{1}{2w}\sum_{i,j}\left({w_{i,j} -\frac{w_i w_j}{2w}}\right)\delta(C_i, C_j)
\label{eq:mod}
\end{align}
where $w_{i,j}$ is the accumulated weight of the arcs between nodes \verb|#i| and \verb|#j|, $w_i$ is the accumulated weight of all arcs of \verb|#i|, $w$ is the total weight 
of the network, $C_i$ is the cluster to which \verb|#i| is assigned, and Kronecker delta $\delta(C_i, C_j)$ is a function, which is equal to $1$ when \verb|#i| and \verb|#j| belong to the same cluster (i.e., $C_i = C_j$), and $0$ otherwise.

Modularity is applicable to non-overlapping cases only. However, there exist modularity extensions that handle overlaps~\cite{Grg11,Chn15}.
The main intuition behind such modularity extensions 
is to quantify the degree of a node membership among multiple clusters either by replacing a Kronecker $\delta$ (see \eqref{eq:mod}) with a similarity measure $s_{ij}$~\cite{Nps08,Shn09} or by integrating a belonging coefficient~\cite{Ncs09,Lzr10,Liu10a} directly into the definition of modularity.
Although both old and new measures are named modularity, they generally have different values even when applied to the same clusters~\cite{Grg08}, resulting in incompatible outcomes. Some modularity extensions are equivalent to the original modularity when applied to non-overlapping clusters~\cite{Nps08,Shn09,Liu10a}. However, the implementations of these extensions introduce an excessive number of additional parameters~\cite{Nps08,Shn09} and/or boost the computational time by orders of magnitude~\cite{Liu10a}, 
which significantly complicates their application to large networks.

\emph{Modularity gain} ($\Delta Q$)~\cite{Bld08} captures the difference in modularity when merging two nodes \verb|#i| and \verb|#j| into the same cluster, providing a computationally efficient way to optimize Modularity:
\begin{align}
\Delta Q_{i,j} &= \frac{1}{2w} \bigg( w_{i,j} - \frac{w_i w_j}{w} \bigg)
\label{eq:dmod}
\end{align}
We use \emph{modularity gain} ($\Delta Q$) as an underlying optimization function for our meta-optimization function MMG (introduced in Section~\ref{subsec:optfunc}).


\paragraph{Degeneracy}\index{degeneracy} is a phenomenon linked to the clustering optimization function appearing when multiple distinct clusterings (i.e., results of the clustering process) share the same globally maximal 
value of the optimization function while being structurally different~\cite{God10}. 
This phenomenon is inherent to any optimization function and implies that a network node might yield the maximal value of the optimization function while being a member of multiple clusters, which is the case when an \textit{overlap} occurs. This prevents the derivation of accurate results by deterministic clustering algorithms without considering overlaps.
To cope with degeneracy, typically multiple stochastic clusterings are produced and combined, 
which is called an \emph{ensemble} or \emph{consensus} clustering and provides robust but coarse-grained results~\cite{God10,Bst13}. 
Degeneracy of the optimization function, together with the aforementioned computational drawback of modularity extensions, motivated us to introduce a new overlap decomposition technique, OD (see Section~\ref{subsec:ovpeval}). OD allows to consider and process overlaps efficiently using algorithms having an optimization function designed for the non-overlapping case. 
It produces accurate, robust and fined-grained results in a deterministic way as we show in our experimental evaluation (see Section~\ref{sec:evals}).

\section{Method}  
\label{sec:method}

We introduce a novel clustering algorithm, \sys, to  perform a stable (i.e., both \emph{robust} and \emph{deterministic}) 
clustering of the input network, producing a fine-grained hierarchy of overlapping clusters.
\sys is a greedy algorithm that uses an agglomerative clustering approach with a local search technique (inspired by Louvain~\cite{Bld08}) and extended with two novel techniques. Namely, we first propose a novel (micro) consensus technique called \emph{Mutual Maximal Gain (MMG)}
for the robust identification of nodes membership in the clusters, which is performed in a deterministic and fine-grained manner.
In addition to MMG, we also propose a new \emph{overlap decomposition (OD) technique} to cope with the degeneracy of the optimization function.  
OD forms stable and fine-grained clusters in a deterministic way from the nodes preselected by MMG.

Algorithm~\ref{alg:cluster} gives a high-level description of our method. 
It takes as input a directed and weighted network with self-loops specifying node weighs.
The resulting hierarchy of clusters is built iteratively starting from the bottom level (the most fine-grained level). 
One level of the hierarchy is generated at each iteration of our clustering algorithm.
A clustering iteration consists of the following steps listed on lines~\ref{aln:identifyCands}--\ref{aln:formClusters}:
\begin{enumerate}
\item Identification of the clustering candidates $ccs_i$ for each node \verb|#i| using the proposed consensus approach, MMG, described in Section~\ref{subsec:mmg} and
\item Cluster formation considering overlaps, described in Section~\ref{subsec:clsform}.
\end{enumerate}
\begin{algorithm}[htbp]\small 
\caption{\sys Clustering Algorithm.}
\label{alg:cluster}
\begin{algorithmic}[1]  
\Function{cluster}{$nodes$}
	\State $hier \gets []$  \Comment List of the hierarchy levels 
	\While{$nodes$} \Comment Stop if the \textit{nodes} list is empty
		\State $\texttt{identifyCands}(nodes)$ \label{aln:identifyCands}  \Comment Initialize \textit{nd.ccs}
		\State $cls \gets \texttt{formClusters}(nodes)$ \label{aln:formClusters}

		\If{$cls$}  \Comment Initialize the next-level nodes  
			\ForAll{$cl \in cls$}
				\State $\texttt{initCluster}(cl)$ \label{aln:initCluster}
			\EndFor

			\ForAll{$nd \in nodes$}  \Comment Consider propagates
				\If{$nd.propagated$}
					\State $\texttt{initNode}(nd)$ \label{aln:initNode}
					\State $cls.append(nd)$
				\EndIf
			\EndFor
			\State $hier.append(cls)$  \Comment Extend the hierarchy  
		\EndIf
		\State $nodes \gets cls$  \Comment Update the processing nodes 
	\EndWhile
	\State \Return $hier$  \Comment{The resulting hierarchy of clusters}
\EndFunction
\end{algorithmic}
\end{algorithm}
At the end of each iteration, links are (re)computed for the formed clusters (\texttt{initCluster} procedure) and for the non-clustered nodes (\emph{propagated} nodes in the \texttt{initNode} procedure).
Both the non-clustered nodes and the formed clusters are treated as input nodes for the following iteration. The algorithm terminates when the iteration does not produce any new cluster.

The clustering process yields a hierarchy of overlapping clusters in a deterministic way independent of the nodes processing order, since all clustering operations
\begin{inparaenum}[\itshape a\upshape)]
\item consist solely 
of non-stochastic, uniform and local operations, and 
\item process each node independently, relying on immutable data evaluated on previous steps only.
\end{inparaenum}
The algorithm is guaranteed to converge since
\begin{inparaenum}[\itshape a\upshape)]
\item the optimization function is bounded (as outlined in Section~\ref{subsec:optfunc})
and monotonically increasing during the clustering process, and
\item the number of formed clusters does not exceed the number of clustered nodes at each iteration (as explained in Section~\ref{subsec:odmarg}).
\end{inparaenum}

\subsection{Identification of the Clustering Candidates}
\label{subsec:clscands}

The \emph{clustering candidates}\index{clustering candidates} are the nodes that are likely to 
be grouped into clusters in the current iteration. 
The clustering candidates are identified for each node ($nd.ccs$) in two steps as listed in Algorithm~\ref{alg:identify_cands}.
First, for each node $nd$ the adjacent nodes ($\{link.dst\; |\, link \in nd.links\}$) having the maximal non-negative value $nd.gmax$ of the optimization function \texttt{optfn} are stored in the $nd.ccs$ sequence, see lines~\ref{aln:cssInitBeg}--\ref{aln:cssInitEnd}.  
Then, the preselected $nd.ccs$ are reduced to the mutual candidates by the \texttt{mcands} procedure, and the filtered out nodes are marked as propagated. The latter step is combined with a cluster formation operation in  our implementation to avoid redundant passes over all nodes. The \texttt{mcands} procedure implements our \textit{Maximal Mutual Gain (MMG)} consensus approach described in Section~\ref{subsec:mmg}, which is a meta-optimization technique that can be applied on top of any optimization function that satisfies a set of constraints described in the following paragraph. 
%
\begin{algorithm}[htbp]\small  
\caption{Clustering Candidates Identification}
\label{alg:identify_cands}
\begin{algorithmic}[1]  
\Function{identifyCands}{$nodes$}
	\ForAll{$nd \in nodes$}  \Comment Evaluate clustering candidates
		\State $nd.gmax \gets -1$ \label{aln:cssInitBeg}  \Comment Maximal gain
		\ForAll{$ln \in nd.links$}
			\State $cgain \gets \texttt{optfn}(nd, ln)$  \Comment Clustering gain
			\If{$cgain \ge 0$ \textbf{and} $cgain \ge nd.gmax$}
				\If{$cgain > nd.gmax$}
					\State $nd.ccs.clear()$  \Comment Reset cands
					\State $nd.gmax \gets cgain$
				\EndIf
				\State $nd.ccs.append(ln.dst)$ \label{aln:cssInitEnd}  \Comment Extend cands
			\EndIf
		\EndFor
	\EndFor

	\Statex
	\ForAll{$nd \in nodes$}  \Comment Reduce the candidates using the consensus approach, propagate remained nodes
		\If{$nd.gmax < 0$ \textbf{or not} \texttt{mcands}$(nd)$}
			\State $nd.propagated \gets \texttt{true}$
		\EndIf
	\EndFor
\EndFunction
\end{algorithmic}
\end{algorithm}

\subsubsection{Optimization Function}
\label{subsec:optfunc}

In order to be used in our method, the optimization function \texttt{optfn} should be
\begin{inparaenum}[\itshape a\upshape)]
\item applicable to pairwise node comparison, i.e. $\exists \, \texttt{optfn}(\verb|#i|, \verb|#j|)\; |\; \verb|#i| \sim \verb|#j|$ (adjusted pair of nodes);
\item commutative, i.e.\linebreak $\texttt{optfn}(\verb|#i|, \verb|#j|) = \texttt{optfn}(\verb|#j|, \verb|#i|)$; and
\item bounded on the non-negative range, 
where positive values indicate some quality improvement in the structure of the forming cluster. 
\end{inparaenum}
There exist various optimization functions satisfying these constraints besides modularity and inverse conductance  
(see the list in~\cite{Csp13}, for instance).

Our \sys algorithm uses \emph{modularity gain}, $\Delta Q$ (see \eqref{eq:dmod}), 
as an optimization function.
We chose modularity (gain) optimization because of the following advantages. First, modularity maximization (under certain conditions) is equivalent to the 
provably 
correct 
but computationally expensive methods of graph partitioning, spectral clustering and to the maximum likelihood method applied to the stochastic block model~\cite{Nwm13,Nmn16}.
Second, there are known and efficient algorithms for modularity maximization, including the Louvain algorithm~\cite{Bld08}, which are accurate and have a near-linear computational complexity.

\subsubsection{MMG Consensus Approach}
\label{subsec:mmg}

We propose a novel (micro) consensus approach, called \emph{Mutual Maximal Gain (MMG)} that
requires only a single pass over the input network, is more efficient and yields much more fine-grained results compared to state-of-the-art techniques. 
\begin{definition}[Mutual Maximal Gain (MMG)]\index{daoc!mutual maximal gain}
\label{dfn:mmg}
\emph{MMG}\index{daoc!MMG} is a value of the optimization function (in our case modularity gain) for two adjacent 
nodes \verb|#i| and \verb|#j|, and is defined in cases where these 
nodes mutually reach the 
maximal value of the optimization function (i.e., reach consensus on the maximal value) when considering each other:
\begin{equation}
\label{eq:mmg}
MMG_{i,j} = \{\Delta Q_{i,j} \;|\; \Delta Q_i = \Delta Q_j,\: \verb|#j| \sim \verb|#i|\}
\end{equation}
where $\sim$ denotes adjacency of \verb|#i| and \verb|#j|, and $\Delta Q_i$ is the maximal modularity gain for \verb|#i|:
\begin{equation}
\label{eq:dmodi}
\Delta Q_i = \{\max{\Delta Q_{i,j}} \;|\; \verb|#j| \sim \verb|#i|\}
\end{equation}
where $\Delta Q_{i,j}$ is the modularity gain for \verb|#i| and \verb|#j| (see \eqref{eq:dmod}). 
\end{definition}

MMG exists in any finite 
network, which can be easily proven by contradiction as follows. 
The nonexistence of MMG would create a cycle with increasing $\max{\Delta Q}$ and results in $\Delta Q_i < \Delta Q_i$ considering that $\forall \verb|#i|\; \exists \Delta Q_i$: $(\Delta Q_i = \Delta Q_{i,j}) < (\Delta Q_j = \Delta Q_{j, k}) < ... < (\Delta Q_t = \Delta Q_{t,j} = \Delta Q_j)$, i.e. $\Delta Q_j < \Delta Q_j$, which yields a contradiction.
MMG evaluation is deterministic 
and 
the resulting nodes are \emph{quasi-uniform} clustering candidates, in the sense that inside each connected component they share the same maximal value of modularity gain. 
MMG takes into account fine-grained clusters as it operates on pairs of nodes, unlike conventional consensus approaches, where micro-clusters either require lots of reexecutions of the consensus algorithm, or cannot be captured at all.
MMG does not always guarantee optimal clustering results but reduces 
degeneracy due to the applied consensus approach.
According to \eqref{eq:mmg}, 
all nodes having MMG to \verb|#i| have the same value of $\Delta Q_i$, i.e., form the overlap in \verb|#i|.
Overlaps processing
is discussed in the following section.

\subsection{Clusters Formation with Overlap Decomposition}  
\label{subsec:clsform}

Clusters are formed from candidate nodes selected by MMG as listed in Algorithm~\ref{alg:form_cls}:
\begin{inparaenum}[\itshape a\upshape)]
\item nodes having a single mutual clustering candidate (\emph{cc}) form the respective cluster directly as shown on line~\ref{aln:clusterSingle},
\item otherwise the overlap is processed. 
\end{inparaenum}
There are three possible ways of clustering an overlapping node in a deterministic way:
\begin{inparaenum}[\itshape a\upshape)]
\item split the node into \emph{fragments} to have one fragment per each \textit{cc} of the node and group each resulting fragment with the respective \textit{cc} into the dedicated cluster (see lines~\ref{aln:splitBeg}--\ref{aln:splitEnd}), or
\item group the node together with all its \textit{nd.ccs} items into a single cluster (i.e. coarsening on line~\ref{aln:clusterAll}), or
\item propagate the node to be processed on the following clustering iteration if its clustering would yield a negative value of the optimization function.
\end{inparaenum}
Each node fragment created by the overlap decomposition is treated as a virtual node representing the belonging degree (i.e., the fuzzy relation) of the original node to multiple clusters. Virtual nodes are used to avoid the introduction of the fuzzy relation for all network nodes (i.e., to avoid an additional complex node attribute) reducing memory consumption and execution time, 
and not affecting the input network itself.
In order to get the most effective clustering result, 
we evaluate the first two aforementioned options and select the one maximizing the optimization function, $\Delta Q$.
Then, we form the cluster(s) by the \texttt{merge} or \texttt{mergeOvp} procedures as follows. The \texttt{mergeOvp} procedure groups each node fragment (i.e., the virtual node created by the overlap decomposition) together with its respective mutual clustering candidate. This results in either
\begin{inparaenum}[\itshape a\upshape)]
\item an extension of the existing cluster to which the candidate already belongs to, 
or
\item the creation of a new cluster and its addition to the \textit{cls} list.
\end{inparaenum}
The \texttt{merge} procedure groups the node with all its clustering candidates either
\begin{inparaenum}[\itshape a\upshape)]
\item merging together the respective 
clusters of the candidates if they exists, or
\item creating a new cluster and adding it to the \textit{cls} list.
\end{inparaenum} 
%
\begin{algorithm}[htbp]\small  
\caption{Clusters Formation}
\label{alg:form_cls}
\begin{algorithmic}[1]
\Function{formClusters}{$nodes$}
	\State $cls \gets []$  \Comment List of the formed clusters
	\ForAll{$nd \in nodes$}
		\If{$nd.propagated$}  \Comment Prefileter nodes
			\State \textbf{continue}
		\EndIf
		
		\Statex
		\If{$nd.ccs.size = 1$}  \Comment Form a cluster
			\State $\texttt{merge}(cls, nd, nd.ccs)$ \label{aln:clusterSingle} 
		\ElsIf{$\texttt{odAccept}(nd)$  
\textbf{and} $\texttt{gainEach}(nd) > \texttt{gainAll}(nd)$} \label{aln:condovp}
			\Comment Form overlapping clusters 
			\ForAll{$cand \in nd.ccs$} \label{aln:splitBeg}
				\State $\texttt{mergeOvp}(cls, nd, cand)$ \label{aln:splitEnd}  
			\EndFor
		\Else  \Comment DBSCAN inspired aggregation
			\State $rccs \gets \texttt{maxIntersectOrig(nd)}$ \label{aln:dbscagg}
			\If{$rccs.size \ge 1$}  \Comment Form a single cluster
				\State $\texttt{merge}(cls, nd, rccs)$  
			\ElsIf{$\texttt{gainAll}(nd) \geq 0$}  \Comment Form a cluster
				\State $\texttt{merge}(cls, nd, nd.ccs)$ \label{aln:clusterAll} \label{aln:condall}
			\Else
				\State $nd.propagated \gets \texttt{true}$
			\EndIf
		\EndIf
	\EndFor
	\State \Return $cls$  \Comment{Resulting clusters}
\EndFunction
\end{algorithmic}
\end{algorithm}

Node splitting is the most challenging process, which is performed only if the accumulated gain from the decomposed node fragments to each of the respective mutual clustering candidates, $nd.ccs$, (\texttt{gainEach} procedure) exceeds the gain of grouping the whole node with all $nd.ccs$ (\texttt{gainAll} procedure). The node splitting involves:
\begin{inparaenum}[\itshape a\upshape)]
\item the estimation of the node fragmentation impact on the clustering convergence (\texttt{odAccept} procedure given in Section~\ref{subsec:odmarg}) and
\item the evaluation of the weights for both the forming fragment and for the links between the fragments of the splitting node as described in Section~\ref{subsec:ovpeval}.
\end{inparaenum}

\subsubsection{Overlap Decomposition (OD)}
\label{subsec:ovpeval}

An overlap occurs when a node has multiple mutual clustering candidates (\emph{ccs}).
To evaluate $\Delta Q$ when clustering the node with each of its $K$ \textit{ccs}, the node is split into $K$ identical and fully interconnected fragments sharing the node weight and original node links. However, since the objective of the clustering is the maximization of $\Delta Q$:
\begin{inparaenum}[\itshape a\upshape)]
\item the node splitting itself is acceptable only in case the resulting $\Delta Q \ge 0$, and
\item the decomposed node can be composed back from the fragments only in case $\Delta Q \le 0$.
\end{inparaenum}
Hence, to have a reversible decomposition without affecting the value of the optimization function for the decomposing node, we end up with $\Delta Q = 0$.

The outlined constraints for an isolated node, which does not have any link to other nodes, can formally be expressed as:
\begin{equation}
\label{eq:ovpres_node}
\begin{cases} 
w = \sum_{k=1}^K w_k + \sum_{k=1}^K \sum_{t=1}^{K-1} \frac{w_{k, t}}{2}\\
\sum_{k=1}^K w_k - \sum_{k=1}^K \frac{(w_k + \sum_{t=1}^{K-1} \frac{w_{k,t}}{2})^2}{w} = 0
\end{cases},
\end{equation}
where $w$ is the weight of the original node being decomposed into $K$ fragments, $w_k$ is the weight of each node fragment $k \in K$ and $w_{k,t}$ is the weight of each link between the fragments. $\Delta Q = Q_{split} - Q_{node} = Q_{split}$ since the modularity of the isolated node is zero (see \eqref{eq:mod}). 
The solution of 
\eqref{eq:ovpres_node} is:
\begin{equation}
\label{eq:ovpres_snode}
w_k = \frac{w}{K^2},\; w_{k,t} = 2\frac{w}{K^2}.
\end{equation}

Nodes in the network typically have 
links, which get split equally between the fragments of the node: 
\begin{equation}
\label{eq:ovpres_sovp}
\forall\: k, t \in \{1\,..\,K\} \;|\; \textit{\#}j \sim \textit{\#}i:\;
\begin{cases}
w_{ik} = \frac{w_i}{K^2}\\ 
w_{ik,it} = 2\frac{w_i}{K^2}\\
w_{ik,j} = 
\frac{w_{i,j}}{K} 
\end{cases},
\end{equation}
where $w_{ik}$ is the weight of each fragment \verb|#ik| of the node \verb|#i|, 
$w_{i,j}$ is the weight of the link $\{\verb|#i|, \verb|#j|\}$. 

\begin{example}[Overlap Decomposition]\index{daoc!overlap decomposition example}
\begin{figure}[bp]\centering  
\includegraphics[scale=0.5]{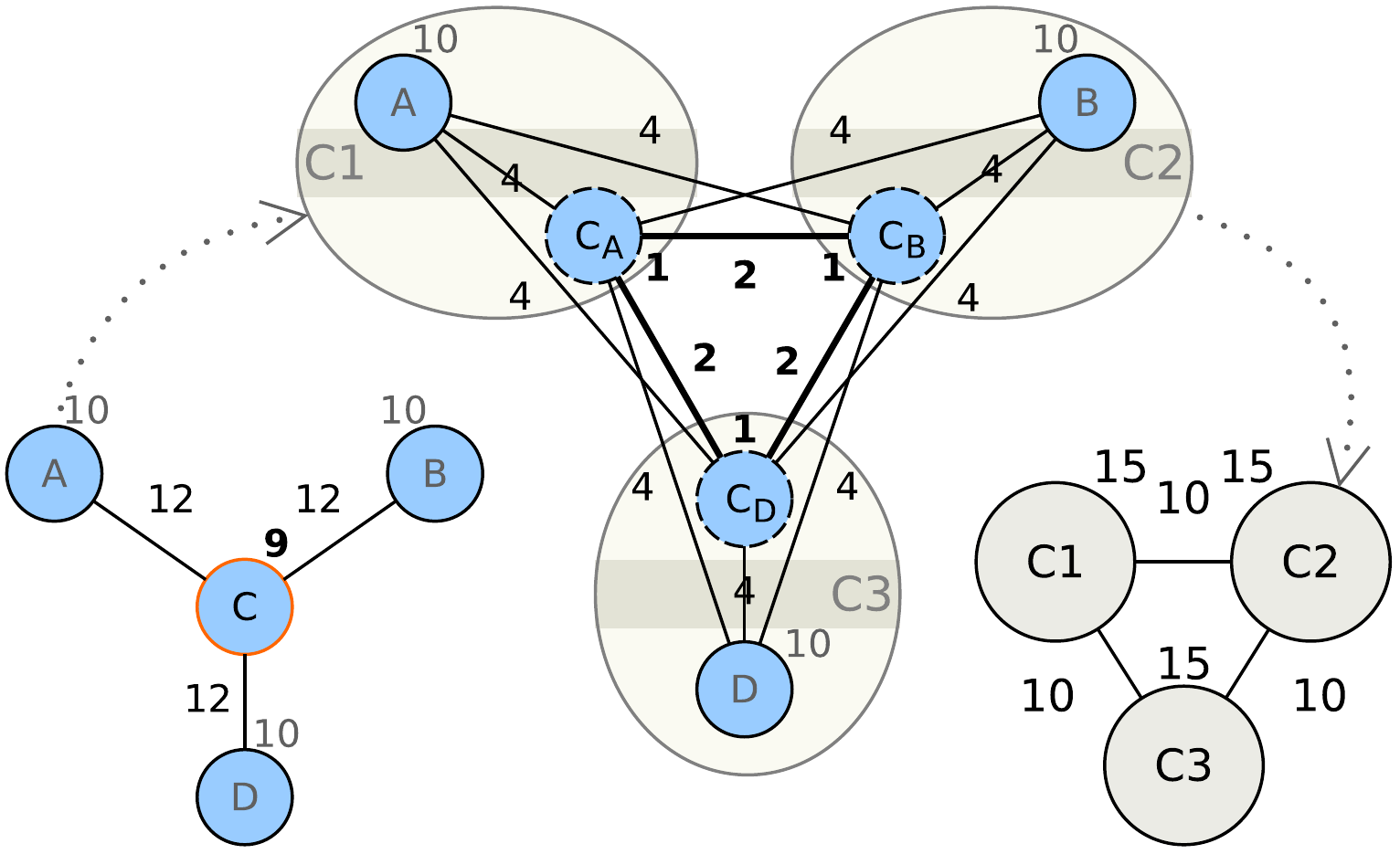}
\caption{Decomposition of the clusters $C1(A,C)$, $C2(B,C)$, $C3(D,C)$ overlapping in node $C$ of the input network $\{A, B, C, D\}$ with weights on both nodes and links.
}
\label{fig:ovpres}
\end{figure}
The input network on the left-hand side of Fig.~\ref{fig:ovpres} has node $C$ with neighbor nodes $\{A, B, D\}$ being \textit{ccs} of $C$. These neighbor nodes form the respective clusters $\{C1, C2, C3\}$ overlapping in $C$. $C$ is decomposed into fragments $\{C_A, C_B, C_D\}$ to evaluate the overlap. 
Node $C$ has an internal weight equal to $9$ (which can be represented via an additional edge to itself) and three edges of weight $12$ each.
The overlapping clusters are evaluated using \eqref{eq:ovpres_sovp} 
as equivalent and virtual non-overlapping clusters formed using the new fragments of the overlapping node:
\begin{align*}
&\forall\: k, t \in \{1\,..\,K\} \;|\; \textit{\#}j \sim \textit{\#}i:\;
\begin{cases}
w_{ik} = \frac{9}{3^2} = 1\\
w_{ik,it} = 2\frac{9}{3^2} = 2\\
w_{ik,j} = 2 + \frac{12}{3} = 6 \quad (\forall j \neq k)\\
w_{ik,k} = \frac{12}{3} = 4
\end{cases}\\
&w_{C_A} = w_{C_B} = w_{C_D} = w_{ik} = 1,\\
&w_{C1} = w_{C2} = w_{C3} = w_k + w_{ik} + w_{ik, k} = 10 + 1 + 4 = 15,\\
&w_{C_k, C_t} = w_{ik,t} + w_{it,k} - w_{ik,it} = 6 + 6 - 2 = 10.
\end{align*}
\label{exm:ovpres}
\end{example}


\subsubsection{Constraining Overlap Decomposition}
\label{subsec:odmarg}

Overlap decomposition (OD) does not affect the value of the optimization function for the node being decomposed ($\Delta Q = 0$), hence it does not affect the convergence of the optimization function during the clustering.
%
However, OD increases the complexity of the clustering when the number of produced clusters exceeds the number of clustered nodes decomposed into multiple fragments. This complexity increase should be identified and prevented to avoid 
indefinite increases in terms of clustering time.

In what follows, we infer a formal condition that guarantees the non-increasing complexity of OD. We decompose a node of degree $d$ into $k$ fragments,
$2 \leq k \leq d$.
Each forming cluster that has an overlap in this node owns one fragment (see Fig.~\ref{fig:ovpres}) and shares at most $d-k$ links to the non-\textit{ccs} neighbors of the node. The number of links between the $k$ fragments resulting in the node split is $k\times \frac{k-1}{2}$.
The aggregated number of resulting links should not exceed the degree of the node being decomposed to retain the same network complexity, therefore:
\begin{equation}
\label{eq:odmargin}
\begin{cases}
k (d - k) + k\frac{k-1}{2} \leq d\\
2 \leq k \leq d
\end{cases}
\end{equation}
The solution of \eqref{eq:odmargin} 
is $2 \leq k \leq d \leq 3$,
namely: $k = 2,\, d = \{2, 3\};\; k = 3,\, d = 3$.

If a node being decomposed has a degree $d >= 3$ or a node has more than $k$ \textit{ccs} then, before falling back to the coarse cluster formation, we apply the following heuristic inspired by the DBSCAN algorithm~\cite{Est96}. 
We evaluate the intersection of \textit{nd.ccs} with each $\{c.ccs\; |\; c \in nd.ccs\}$ (\texttt{maxIntersectOrig} procedure on line~\ref{aln:dbscagg} of Algorithm~\ref{alg:form_cls}) and group the node with its clustering candidate(s) yielding the maximal 
intersection if the latter contains at least half of the \textit{nd.ccs}. In such a way, we try prevent an early 
coarsening and obtain more fine-grained and accurate results.

\subsection{Complexity Analysis}  
\label{subsec:complexity}

The computational complexity of \sys on sparse networks is $O(m \cdot \log{m})$, where \textit{m} is the number of links in the network. 
All links of each node ($d \cdot n = m$) are processed for $\log{m}$ iterations. In the worst case, the number of iterations is equal to the number of nodes \textit{n} (instead of $\log{m}$) and the number of mutual candidates is equal to the node degree \textit{d} instead of \textit{1}. Thus, the theoretical worst-case complexity is $O(m \cdot d \cdot n) = O(m^2)$ and occurs only in a hypothetical dense symmetric network having equal MMG for all links (and, hence, requiring overlap decomposition) on 
each clustering iteration and in case the number of clusters is decreased at each iteration by one only.
The memory complexity is $O(m)$. 

\section{Experimental Evaluation}  
\label{sec:evals}

\subsection{Evaluation Environment}
\label{subsec:evalenv}

Our evaluation was performed using an open-source parallel isolation benchmarking framework, \bmark\footnote{\urlBmark\label{ftn:bmark}}~\cite{Clb18}, on a Linux Ubuntu 16.04.3 LTS server with the Intel Xeon CPU E5-2620 v4 @ 2.10GHz CPU (16 physical cores) and 132 GB RAM.
The execution termination constraints for each algorithm are as follows:
64 GB of RAM and 72 hours max per network clustering. 
Each algorithm is executed on a single dedicated physical CPU core with up to 64 GB of guaranteed available physical memory.

\begin{table*}[bp]
\centering
\caption{Evaluating clustering algorithms.}
\label{tbl:algs}
\vspace{-8pt}
\rowcolors{2}{gray!25}{white}
\begin{tabular}{@{}lccccccccccc@{}}
\toprule
\textbf{Features \textbackslash\, Algs}
& \textbf{\textit{Daoc}}
& \textbf{Scp} & \textbf{Lvn} & \textbf{Fcl} & \textbf{Osl2} & \textbf{Gnx} & \textbf{Psc} & \textbf{Cgr} & \textbf{Cgri} & \textbf{Scd} & \textbf{Rnd}
\\ \midrule
Hierarchical                          & +             &              & +                &                  & +               &                 &                &                   &                    &              &                      \\
Multi-scale                           & +             & +            & +                &                & +               & +               &                &                   &                    &              &                      \\
Deterministic                         & +             & +            &                  &                  &                 &                 & +              &                   &                    &             &                      \\
Overlapping clusters                         & +             & +            &                  &                  & +               & +               & +              &                   &                    &              &                      \\
Weighted links                          & +             & +            & +                & $\circ$               & +               & +               &                &                   &                    &              & +                    \\
Parameter-free                        & +!            &              & +                & *               & *               & *               &                & *                 & *                  & *            & +                    \\
Consensus/Ensemble                    & +             &              &                  & +               & +               &                 &                & +                 & +                  &              &                      \\ \bottomrule
\end{tabular}
\begin{flushleft}
\footnotesize{
\emph{Deterministic} includes input-order invariance;\\
+!  the feature is available, 
still the ability to force custom parameters is provided;\\
*  the feature is partially available, 
parameters tuning might be required for specific cases;\\ 
$\circ$  the feature is available in theory but is not supported by the original implementation of the algorithm.
}
\end{flushleft}
\vspace{-4pt}
\end{table*}
We compare \sys against almost a dozen state-of-the-art clustering algorithms 
listed in Table~\ref{tbl:algs} (the original implementations of all algorithms except Louvain are included into \bmark and are executed as precompiled or JIT-compiled applications or libraries) and described in the following papers: SCP~\cite{Kpl08}, Lvn(Louvain\footnote{\url{http://igraph.org/c/doc/igraph-Community.html}}~\cite{Bld08}), Fcl (Fast Consensus on Louvain: FCoLouv~\cite{Adt19}), Osl2(OSLOM2~\cite{Lcn11}), 
Gnx(GANXiS also known as SLPA~\cite{Xie11}), Psc(pSCAN~\cite{Chn16}), Cgr[i](CGGC[i]\_RG~\cite{Ogn13}), SCD~\cite{Prz14} and Rnd(Randcommuns~\cite{Clb18}). 
We have not evaluated a well known CPM-based overlapping clustering algorithm, CFinder~\cite{Grl05}, because
\begin{inparaenum}[\itshape a\upshape)]
\item GANXiS outperforms CFinder in all aspects by several accuracy metrics~\cite{Xie11,Xie13} and
\item we do evaluate SCP, a fast CPM-based algorithm.
\end{inparaenum}
For a fair accuracy evaluation, we uniformly sample up to 10 levels from the clustering results (levels of the hierarchical / multilevel output or clusterings produced uniformly varying algorithm parameters in the operational range) and take the best value.

\subsection{Stability Evaluation}  
\label{subsec:evaltarg}

We evaluate stability in terms of both robustness and determinism for the consensus (ensemble) and deterministic clustering algorithms listed in Table~\ref{tbl:algs}.
Determinism (non-stochasticity and input order independence) evaluation is performed on synthetic and real-world networks below, where we quantify the standard deviation of the clustering accuracy.
To evaluate stability in terms of robustness, 
we quantify the deviation of the clustering accuracy in response to small perturbations of the input network.
The clustering accuracy on each iteration is measured relative to the clustering yielded by the same algorithm at the previous perturbation iteration. For each clustering algorithm, the accuracy is evaluated only for the middle level (scale or hierarchical level), since it is crucial to take the same clustering scale to quantify structural changes in the forming clusters of evolving networks. Robust clustering algorithms are expected to have their accuracy gradually evolving (without any \emph{surges}) relative to the previous perturbation iteration. In addition, the clustering algorithms sensitive enough to capture the structural changes are expected to have their accuracy \emph{monotonically decreasing} since the relative network reduction (perturbation) is increased at each iteration: $X$ deleted links on iteration $i$ represent a fraction of $X/N_i$, but on the following iteration this fraction is increased to $X/(N_i - X)$.

We evaluate robustness and sensitivity (i.e., the ability to capture small structural changes) on synthetic networks with nodes forming overlapping clusters generated by the LFR framework~\cite{Lcn09b}. We generate a synthetic network with ten thousand nodes having an average degree of 20 and using the mixing parameter $\mu = 0.275$. This network is shuffled (the links and nodes are reordered) 4 times to evaluate the input order dependence of the algorithms. Small perturbations of the input data are performed gradually reducing the number of links in the network by 2\% of the original network size (i.e., 10 $\times$ 1000 $\times$ 20 $\times$ 0.02 = 4000 links) starting from 1\% and ending at 15\%. The links removal is performed
\begin{inparaenum}[\itshape a\upshape)]
\item randomly to retain the original distributions of the network links and their weights but
\item respecting the constraint that each node retains at least a single link.
\end{inparaenum}
This constraint prevents the formation of disconnected regions. Our perturbation does not include any random modification of the link weights or the creation of new links since it would affect the original distributions of the network links and their weights, causing surges in the clustering results.


\begin{figure}[!b]  
\vspace{-12pt}
\centering
\subfigure[F1h (average value and deviation) for subsequent perturbations (link removals). Stable algorithms are expected to have a gracefully decreasing F1h without any surges.]{
\includegraphics[scale=0.6]{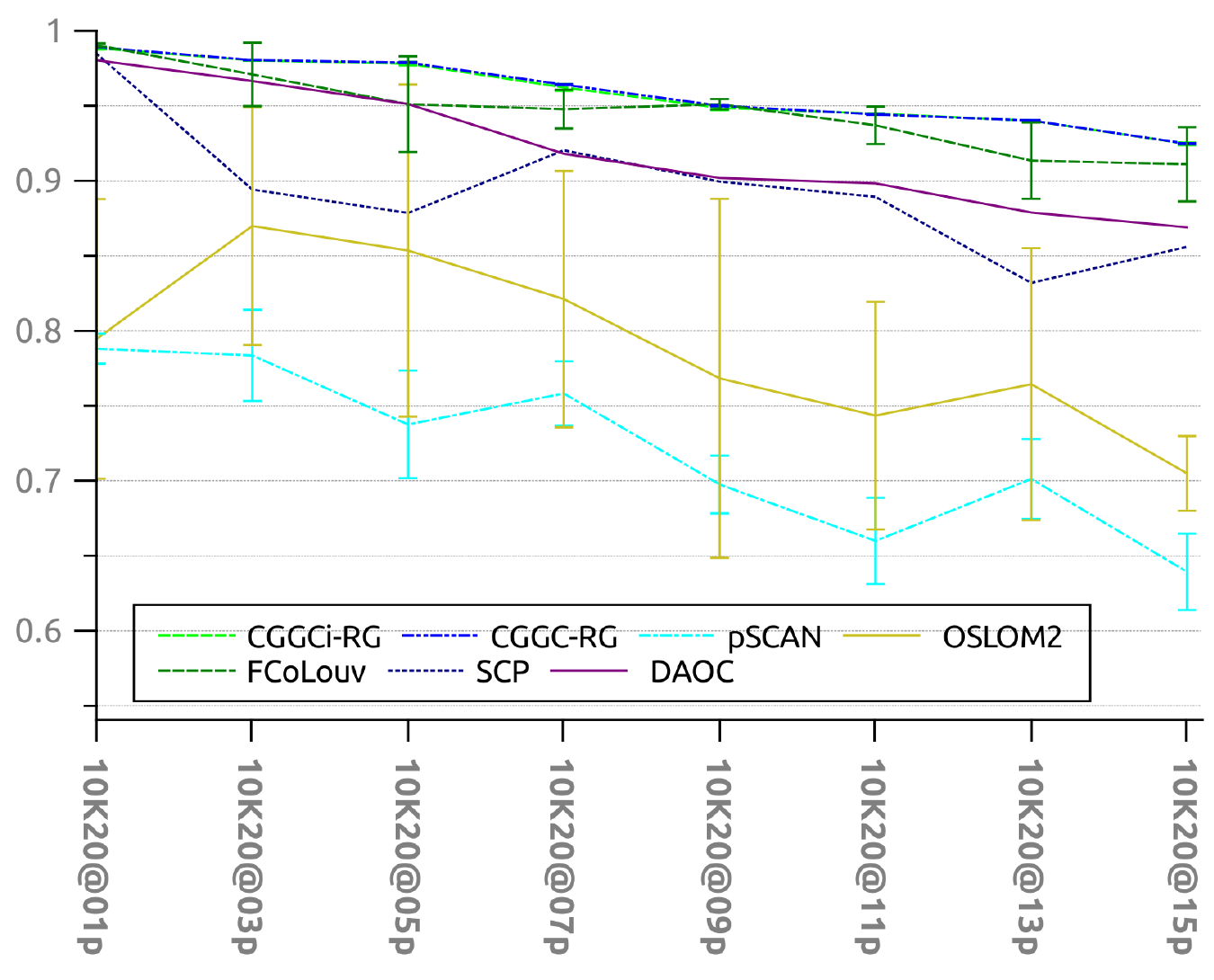}
\label{fig:f1h_lreduct}
}
\quad
\subfigure[$\Delta$F1h relative to the previous perturbation iteration. Stable and sensitive algorithms are highlighted with bolder line width and have positive $\Delta$F1h evolving without surges. Standard deviation is shown only for the consensus algorithms but visible only for FCoLouv and CGGCi-RG.]{
\includegraphics[scale=0.6]{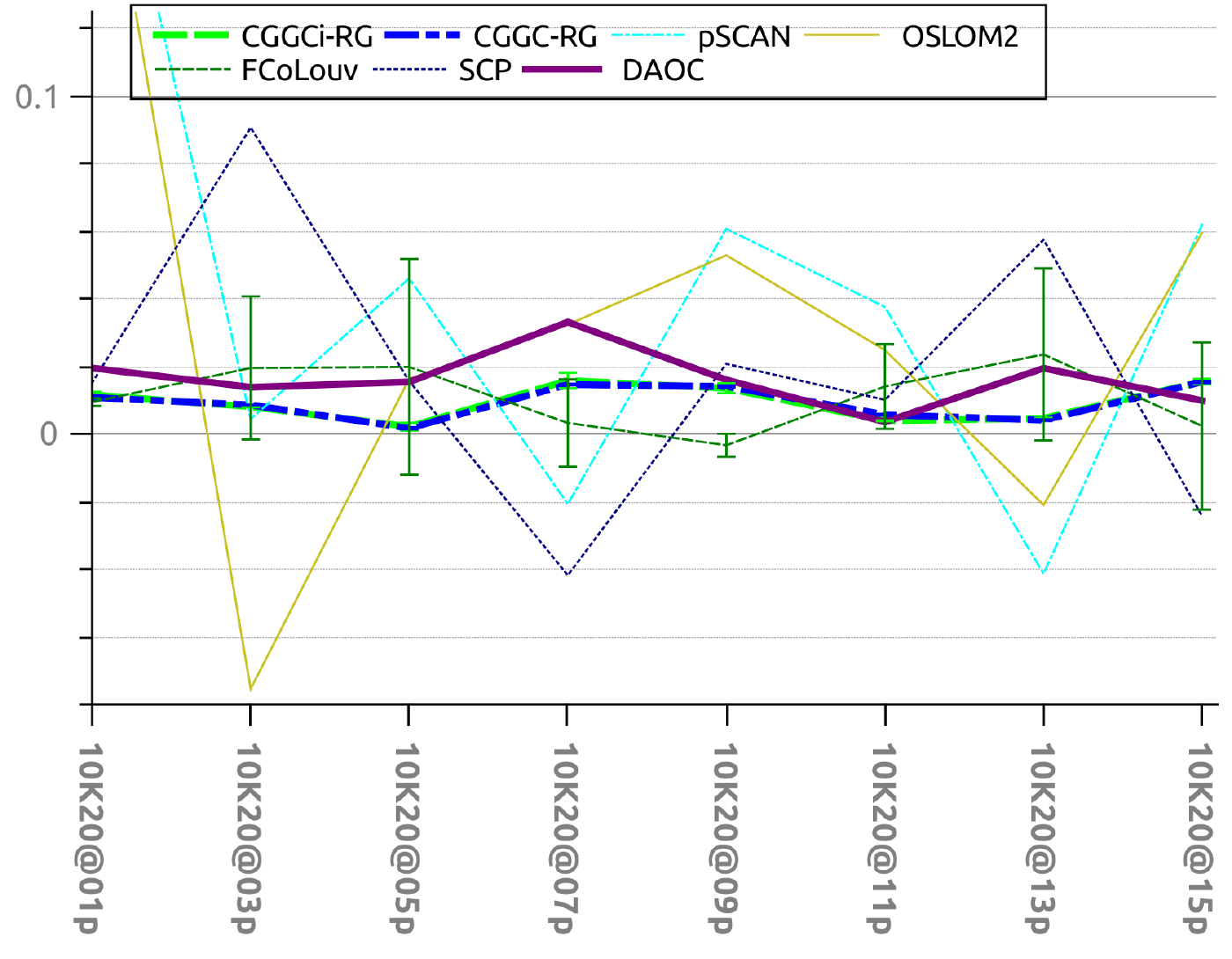}
\label{fig:df1h_lreduct}
}
\vspace{-4pt}
\caption{Stability and sensitivity evaluation.
}
\label{fig:eval_lreduct}
\end{figure}
%
The evaluations of stability in terms of robustness (absence of surges in response to small perturbation of the input network) and sensitivity (ability to capture small structural changes) are shown in Fig.~\ref{fig:eval_lreduct}. Absolute accuracy values relative to the previous link reduction iteration are shown in Fig.~\ref{fig:f1h_lreduct}. The results demonstrate that, as expected, all deterministic clustering algorithms except \sys (i.e. pSCAN and SCP) result in surges and hence are not robust. 
We also obtain some unexpected results. First, pSCAN, 
which is nominally ``exact'' (i.e., non-stochastic and input-order independent), actually shows significant deviations in accuracy. 
Second, the clusterings produced by OSLOM2 using default parameters and by FCoLouv using a number of consensus partitions $n_p = 5$ 
are prone to surges. Hence, OSLOM2 and  FCoLouv cannot be classified as robust algorithms according to the obtained results in spite of being a consensus clustering algorithms.
Fig.~\ref{fig:df1h_lreduct} illustrates the sensitivity of the algorithms, where the relative accuracy values compared to the previous perturbation iteration are shown. Sensitive algorithms have monotonically decreasing results for the subsequent link reduction, which corresponds to positive values on this plot. The stable algorithms (CGGC-RG, CGGCi-RG and \sys) 
are highlighted with a bolder line width on the figure.
These results demonstrate that being robust, CGGC-RG and CGGCi-RG are not always able to capture structural changes in the network, i.e., they are less sensitive than \sys. 
Overall, the results show that only our algorithm, \sys, is stable (it is deterministic, including input-order independence, and robust) and at the same time 
is able to capture even small structural changes in the input network.

\subsection{Effectiveness and Efficiency Evaluation}
\label{subsec:evalcore}

Our performance evaluation was performed both
\begin{inparaenum}[\itshape a\upshape)]
\item on weighted undirected synthetic networks with overlapping ground-truth clusters produced by the LFR
framework
integrated into \bmark~\cite{Clb18} and
\item on large real-world networks having overlapping and nested ground-truth communities\footnote{\url{https://snap.stanford.edu/data/\#communities}}~\cite{Yan15}.
\end{inparaenum}
The synthetic networks were generated with 1, 5, 20 and 50 thousands nodes, each having an average node degrees of 5, 25 and 75. 
The maximal node degree is uniformly scaled from 70 on the smallest networks up to 800 on the largest ones. 
Synthetic networks generation parameters 
are taken by default as provided by \bmark. The real-world networks contain from 334,863 nodes with 925,872 links (amazon) up to 3,997,962 nodes with 34,681,189 links (livejournal). 
The ground-truth communities of real-world networks were pre-processed to exclude duplicated clusters (communities having exactly the same nodes).
Each network is shuffled (reordered) 4 times and the average accuracy value along with its standard deviation are reported.

A number of accuracy measures exist for overlapping clusters evaluation. We are aware of only two families of accuracy measures applicable to large overlapping clusterings, i.e. having a near-linear computational complexity with the number of nodes: the F1 family~\cite{Xms19} and generalized NMI (GNMI)~\cite{Esv12,Xms19}. However, mutual information-based measures are biased to a large numbers of clusters while GNMI does not have any bounded computational complexity in general.
Therefore, we evaluate clustering accuracy with \emph{F1h}
~\cite{Xms19}, a modification of the popular \textit{average F1-score (F1a)}~\cite{Yng13,Prz14} providing indicative values in the range $[0, 0.5]$, since the artificial clusters formed from all combinations of the input nodes yield $F1a \rightarrow 0.5$ and $F1h \rightarrow 0$.

\begin{figure}[tbp] 
\vspace{-8pt}
\includegraphics[scale=0.54]{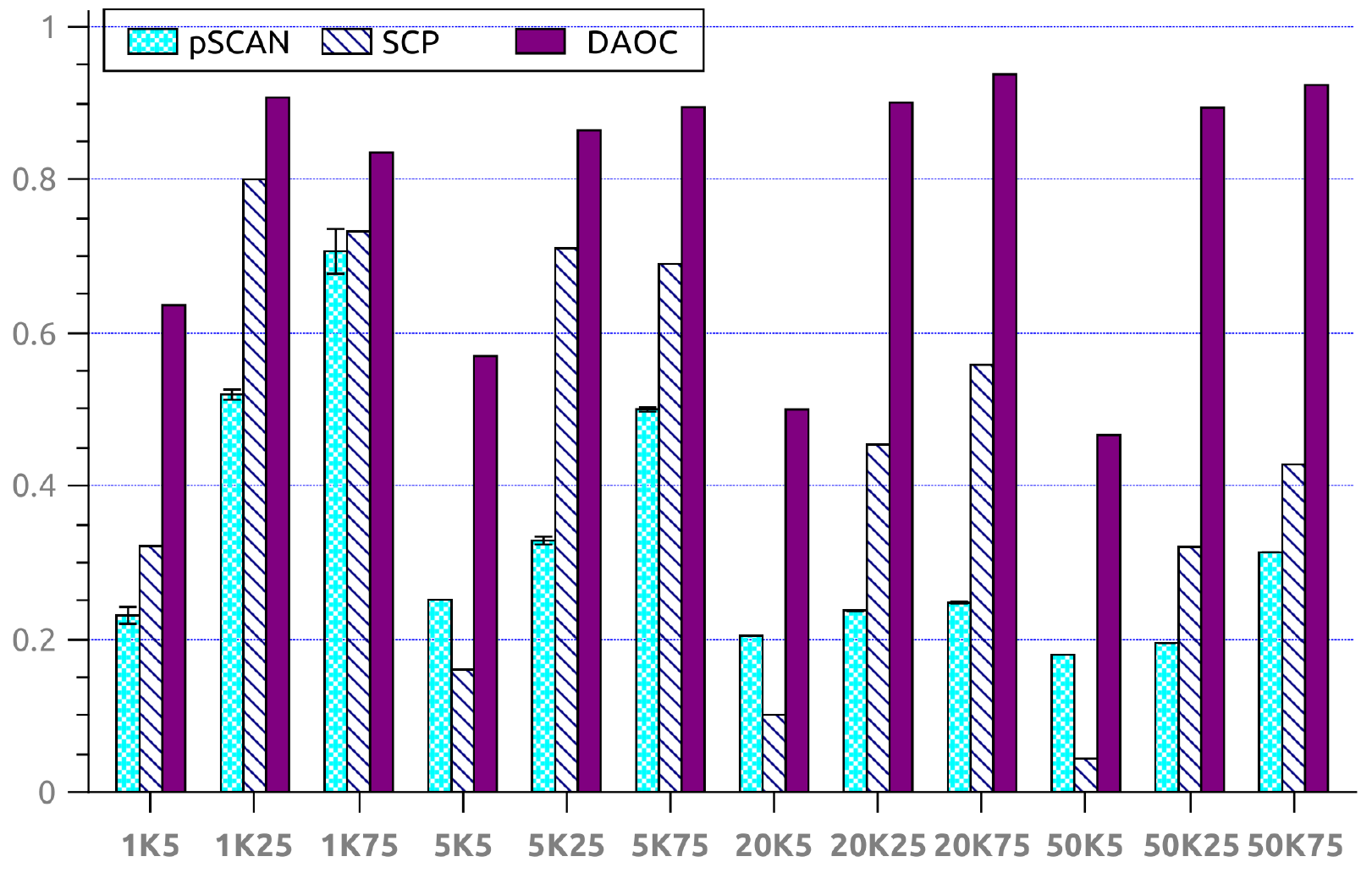}  
\caption{F1h of the deterministic algorithms on the synthetic networks.} \label{fig:f1h_syntnets}
\vspace{-8pt}
\end{figure}

\begin{figure}[htbp]
\vspace{-8pt}
\centering
\subfigure[F1h (average value and deviation).]{
\includegraphics[scale=0.6]{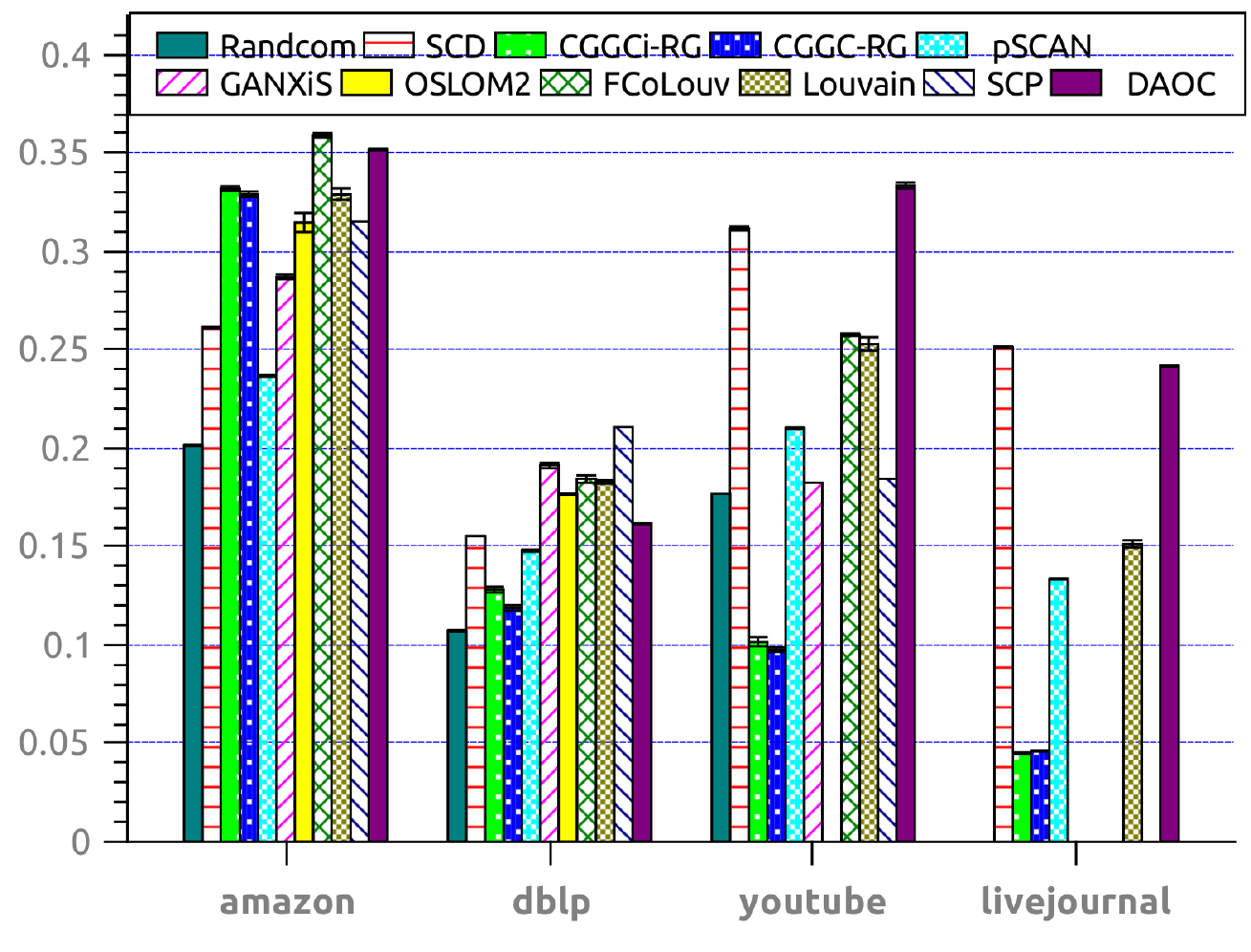}  
\label{fig:f1h_realnets}
}
\quad
\subfigure[Execution time for a single algorithm run 
on a single and dedicated CPU core, sec.
The range in SCP shows the execution time for $k > 3$.
]{
\includegraphics[scale=0.85]{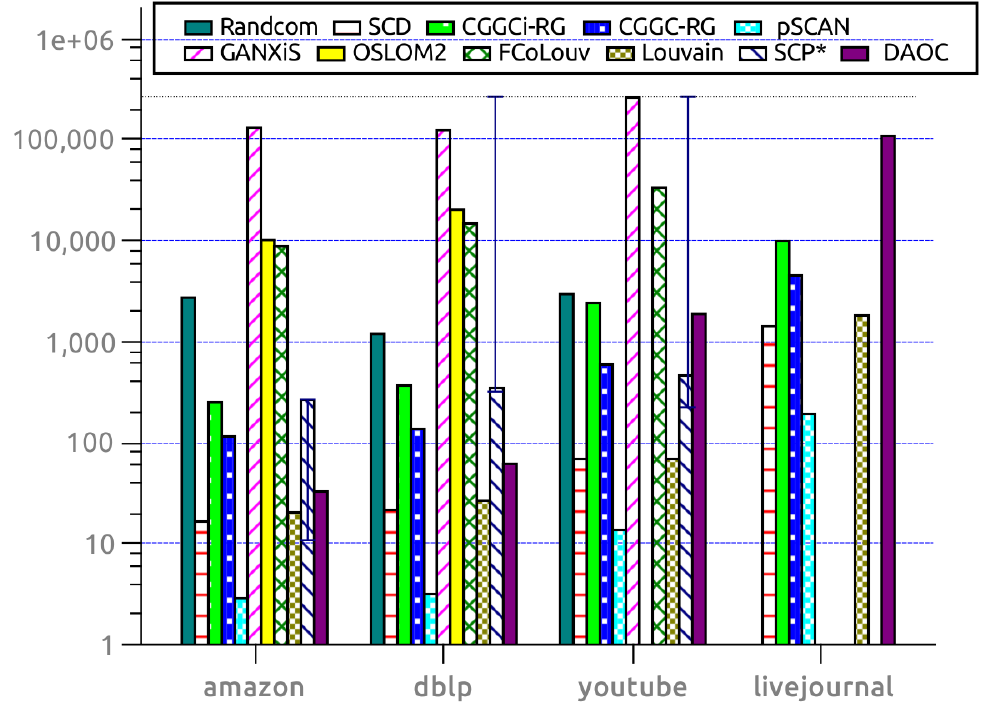}
\label{fig:time_realnets}
}
\vspace{-4pt}
\caption{Performance on the real-world networks.}
\label{fig:eval_realnets}
\vspace{-8pt}
\end{figure}

First, we evaluate accuracy for all the deterministic algorithms listed in Table~\ref{tbl:algs} on synthetic networks, and then evaluate both accuracy and efficiency for all clustering algorithms on real-world networks.
Our algorithm, \sys, shows the best accuracy among the deterministic clustering algorithms on synthetic networks, outperforming others on each network and being more accurate by 25\% on average according to Fig.~\ref{fig:f1h_syntnets}.
Moreover, \sys also has the best accuracy on average among all evaluated algorithms on large real-world networks as shown in Fig.~\ref{fig:f1h_realnets}.
Being parameter-free, our algorithm yields good accuracy on \emph{both} synthetic networks and real-world networks, 
unlike some other algorithms having good performance on some datasets but low performance on others.

%
%
\begin{table*}[htbp]
\centering
\caption{Peak memory consumption (RSS) on the real-world networks, MB.}
\label{tbl:mem}
\vspace{-8pt}
\rowcolors{2}{gray!25}{white}
\begin{tabular}{@{}lccccccccccc@{}}
\toprule
\textbf{Nets\textbackslash Algs} 
& \textbf{\textit{Daoc}}
& \textbf{Scp*} & \textbf{Lvn} & \textbf{Fcl} & \textbf{Osl2} & \textbf{Gnx} & \textbf{Psc} & \textbf{Cgr} & \textbf{Cgri} & \textbf{Scd} & \textbf{Rnd}
\\ \midrule
amazon                          & 238             & 3,237            & 339                & 3,177                & 681               & 3,005                & 155               & 247                  & 1,055                   & \emph{37}             & 337                     \\
dblp                           & 225             & 3,909            & 373                & 3,435                & 717               & 2,879               & 167               & 247                  & 1,394                   & \emph{36}             & 373                     \\
youtube                         & 737             & 4,815            & 1,052                 & --                & --                & 8,350                & 508              & 830                  & 3,865                   & \emph{131}            & 1,050                     \\
livejournal                         & 5,038             & --            & 10,939                 & --               & --               & --               & 4,496              & 4,899                  & 11,037                   & \emph{761}             & --                     \\ \bottomrule
\end{tabular}
\begin{flushleft}
\footnotesize{
--  denotes that the algorithm was terminated for violating the execution constraints;\\
*  the memory consumption and execution time for SCP are reported for a clique size $k=3$ 
since they grow exponentially with $k$ on dense networks, though accuracy was evaluated varying $k \in 3 .. 7$. 
}
\end{flushleft}
\vspace{-4pt}
\end{table*}

Besides being accurate, \sys consumes the least amount of memory among the evaluated hierarchical algorithms (Louvain, OSLOM2) as shown in Table~\ref{tbl:mem}. In particular, \sys consumes 2x less memory than Louvain on the largest real-world evaluated network (livejournal) and 3x less memory than OSLOM2 on dblp, while producing much more fine-grained hierarchies of clusters with almost an order of magnitude more levels than other algorithms. Moreover, among the evaluated overlapping clustering algorithms, only pSCAN and \sys are able to cluster the livejournal network within the specified execution constraints, the missing bars in Fig.~\ref{fig:time_realnets} corresponding to the algorithms that we had to terminate.



%
%


\section{Conclusions}

In this paper, we presented a new clustering algorithm, \sys, which is at the same time stable and provides a unique combination of features yielding a fine-grained hierarchy of overlapping clusters in a fully automatic manner. 
We experimentally compared our approach on a number of different datasets and showed that while being parameter-free and efficient, it yields accurate and stable results on \emph{any} input networks. \sys builds on a new (micro)  
consensus technique, MMG,
and a novel overlap decomposition approach, OD, which are both applicable on top of non-overlapping clustering algorithms and allow to produce overlapping and robust clusters. 
\sys is released as an open-source clustering library implemented in C++ that  
includes various cluster analysis features not mentioned in this paper and that 
is integrated with several data mining applications (\citeStx, or DAOR
~\cite{Daor19} embeddings).  
In future work, we plan to design an approximate version of MMG to obtain near-linear execution times on dense networks, and to parallelize \sys taking advantage of modern hardware architectures to further expand the applicability of our method.



%
%

\bibliographystyle{IEEEtran}\balance
\bibliography{IEEEabrv,./daoc}

\end{document}